\documentclass[journal,twocolumn,]{IEEEtran}
\usepackage{amsmath}
\usepackage{graphicx}
\usepackage[utf8]{inputenc}
\usepackage{siunitx}
\usepackage{upgreek}
\usepackage{stackengine}
\usepackage{algorithmic}
\usepackage{algorithm,algorithmic} 
\usepackage{amssymb}
\usepackage{optidef}
\usepackage{xcolor}
\usepackage{enumitem}
\usepackage{authblk}
\usepackage{cuted}
\newlist{steps}{enumerate}{1}
\setlist[steps, 1]{label = \textbf{Step} \arabic*:}
\usepackage{cite}
\usepackage[nolist]{acronym}
\usepackage{pgfplotstable}
\usepgfplotslibrary{patchplots}
\usepackage{authblk}

\begin{document}
\def\Ns{  {N}_{\textrm{s}}  }
\def\Nr{  {N}_{\text{r}}  }
\def\x{x_{\tiny{\oplus}}}
\def\xtilde{\widetilde{\mathbf{x}}}
\def\xrelay{\widehat{x}_{\tiny{\oplus}\mbox{,}\mathrm{r}}}
\def\xDone{\widehat{x}_{\tiny{\oplus}\mbox{,}\mathrm{D}_{1}}}
\def\xDtwo{\widehat{x}_{\tiny{\oplus}\mbox{,}\mathrm{D}_{2}}}
\def\BERrelay{\mathbb{P}^{\mathrm{r}}_{\tiny{\oplus}}}
\def\BERDone{\mathbb{P}^{\mathrm{D}_{1}}_{x_{2}}}
\def\BERSD{\mathbb{P}^{\mathrm{S}_{1}}_{\mathrm{D}_{1}}}
\def\BERRD{\mathbb{P}^{\mathrm{r}}_{\mathrm{D}_{1}}}
\def\BERD{\mathbb{P}_{\mathrm{D}_{1}}}
\begin{acronym}
\acro{5G}{fifth-generation}
\acro{BER}{bit error rate}
\acro{BPSK}{binary phase shift keying}
\acro{CCP}{concave convex procedure}
\acro{CSI}{channel state information}
\acro{DC}{difference-of-convex functions}

\acro{IoT}{Internet of things}
\acro{IRS}{intelligent reflecting surface}

\acro{LLR}{log likelihood ratio}

\acro{MIMO}{multiple-input multiple-output}
\acro{mMTC}{massive machine type communication}
\acro{mm-Wave}{millimeter wave}
\acro{MMSE}{minimum mean square error}
\acro{MSE}{mean-squared error}

\acro{NNC}{network-layer network coding}

\acro{OAC}{over-air-computation}

\acro{pdf}{probability density function}
\acro{PNC}{physical layer network coding}
\acro{QPSK}{quadrature phase shift keying}

\acro{SNR}{signal to noise ratio}

\end{acronym}
	\pgfplotsset{every axis/.append style={
			line width=1pt,
			legend style={font=\large, at={(0.97,0.85)}}},
	} %
	\pgfplotsset{compat=1.16}
\title{Wireless Network Coding with Intelligent Reflecting Surfaces}
\author{Amanat Kafizov,~\IEEEmembership{Student Member,~IEEE}, Ahmed Elzanaty,~\IEEEmembership{Member,~IEEE}, Lav R. Varshney,~\IEEEmembership{Senior Member,~IEEE}, Mohamed-Slim Alouini,~\IEEEmembership{Fellow,~IEEE}
\thanks{A. Kafizov, A. Elzanaty, and  M.-S. Alouini are with Computer, Electrical, and Mathematical Science and Engineering (CEMSE) Division, King Abdullah University of Science and Technology (KAUST), Thuwal, Saudi Arabia (e-mail: \{amanat.kafizov,ahmed.elzanaty, slim.alouini\}@kaust.edu.sa).}
\thanks{L. R. Varshney is with the Coordinated Science Laboratory and the Department of Electrical and Computer Engineering, University of Illinois at Urbana-Champaign, Urbana, IL, USA (e-mail: varshney@illinois.edu).}
}

\maketitle

\begin{abstract}
    Conventional wireless techniques are becoming inadequate for beyond \ac{5G} networks due to latency and bandwidth considerations.
    To improve the error performance and throughput of wireless communication systems, we propose \ac{PNC} in an \ac{IRS}-assisted environment.  We consider an  \ac{IRS}-aided butterfly network, where we propose an algorithm for obtaining the optimal \ac{IRS} phases. Also, analytic expressions for the \ac{BER} are derived. The numerical results demonstrate that the proposed scheme significantly improves the \ac{BER} performance. For instance, the \ac{BER} at the relay in the presence of a $32$-element \ac{IRS} is three orders of magnitudes less than that without an \ac{IRS}. 
\end{abstract}

\begin{IEEEkeywords} 
		Intelligent reflecting surfaces; network coding; butterfly networks; performance analysis
\end{IEEEkeywords}
	\acresetall 

\section{Introduction}
Conventional wireless communication techniques are becoming inadequate for beyond \ac{5G} networks because they, in general, have excessive network latency and low spectral efficiency \cite{guo2017massive}.
\emph{Network coding} is a promising solution to increase the throughput of  wireless networks \cite{katti2008xors,sykora2018wireless}. It relies on the ability of the relay (router) to perform more complex operations than just forwarding \cite{kosut2009nonlinear}. One kind of network coding is \ac{PNC}, where the physical broadcast nature of wireless links, which generally causes deleterious interference, is exploited to increase the network throughput \cite{zhang2010physical,katti2008xors,sykora2018wireless,kosut2009nonlinear,lee2010degrees}.
%
Yet, higher bandwidth is required to support \ac{PNC} with higher data rates, which necessitates a move toward the \ac{mm-Wave} band. However, \ac{mm-Wave} communication suffers from high path loss and is susceptible to blockages. These limitations suggest future sustainable networks, where the propagation channel itself can be controlled. 


In this regard, so-called \acp{IRS} have been demonstrated to overcome some of the issues associated with \acp{mm-Wave} \cite{di2019smart}. 
An \ac{IRS} is a flat surface which comprises many small passive elements, each of which can independently introduce phase changes to the incident signals \cite{wu2020intelligent}. Thus, unlike traditional relays,  it does not need a dedicated energy source or consume any transmit power. An \ac{IRS} can be easily integrated into walls, ceilings, and building facades \cite{wu2019towards}. \acp{IRS} can enhance communication performance in terms of coverage, energy efficiency, electromagnetic radiation reduction, and wireless localization accuracy \cite{KiskSlim:20,huang2019reconfigurable,IbrElzanaty:21,ElzChiSlim:20,ElzanatyUeGuiSlim:20}. Also, \acp{IRS} can improve the \ac{BER} performance of \ac{OAC} techniques \cite{fang2020stochastic,jiang2019over,yu2020optimizing}.  This is achieved by smartly tuning the \ac{IRS} phase profile, thereby significantly boosting the received signal power. In a sense, \ac{PNC} is a kind of \ac{OAC} as it is based on the linear or nonlinear aggregation of signals by the wireless medium; yet it has remained unstudied in the presence of \acp{IRS}.


 In this paper, we propose a \ac{PNC} scheme in an \ac{IRS}-assisted environment to enhance the error performance and throughput of wireless communication systems, 
 especially in unfavorable channel conditions for \ac{mm-Wave} communication. As a proof of concept, we focus on the butterfly network, usually adopted as the basic setting in the framework of network coding \cite{zhang2010physical,ahlswede2000network}. In this setting, the relay computes network-coded packets directly from the received signal. To improve \ac{BER} performance, we optimize the \ac{IRS} profile. Although, the problem is non-convex, we propose an efficient optimization algorithm based in matrix lifting. Then, we derive the analytic \ac{BER} of \ac{IRS}-aided butterfly network to judge error performance. Numerical results verify the significant improvement of \ac{BER} performance due to the proposed scheme.

For notation, the probability of an event and expectation of a random variable are denoted by $\mathbb{P}\{\cdot\}$ and $\mathbb{E}\left[\cdot\right]$, respectively. Capital bold letters, e.g., $\mathbf{X}$, and small bold letters, e.g., $\mathbf{x}$, denote matrices and vectors, respectively. The notation $\operatorname{diag}(\mathbf{x})$ represents a diagonal matrix with diagonal $\mathbf{x}$, while $\operatorname{tr}\left(\cdot\right)$ denotes the matrix trace. The transpose and Hermitian of vectors or matrices are denoted as $(\cdot)^{T}$ and $(\cdot)^{H}$. The $l^{th}$ row and column of a matrix $\mathbf{X}$ are denoted by $\mathbf{x}_{(l)}$ and $\mathbf{x}_{l}$, respectively. The real part of complex number $z$ is denoted by $\Re{\left(z\right)}$.  The $\ell_2$-norm of a vector $\mathbf{x}$ is denoted by $\left|\left|\mathbf{x}\right|\right|$.  

\section{System Model}\label{sec_sys_mod}
\begin{figure}
    \centering
    \includegraphics[width=3.4in]{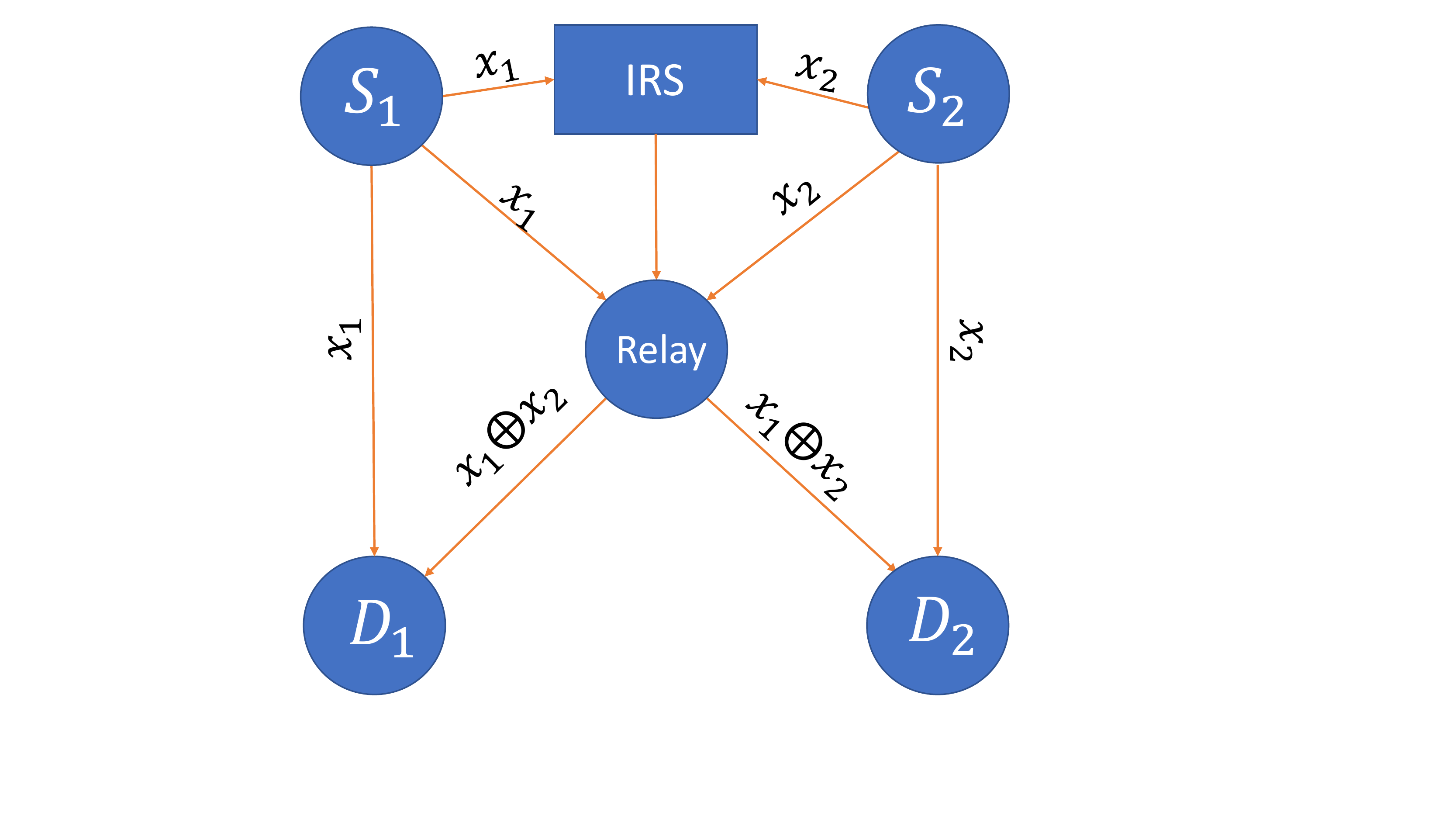}\vspace{-9mm}
    \caption{IRS-aided butterfly network.}
    \vspace{-0.2cm}
    \label{Butterfly_scheme}
\end{figure}
This paper considers the \ac{IRS}-aided butterfly network as shown in Fig. \ref{Butterfly_scheme}, where sources $\mathrm{S_{1}}$ and $\mathrm{S_{2}}$ want to deliver their messages to destinations $\mathrm{D_{1}}$ and $\mathrm{D_{2}}$. For simplicity, we assume each source and destination node have a single antenna to transmit and receive, respectively, whereas the relay node is equipped with $\Nr=2$ antennas. To enhance communication performance from $\Ns=2$ sources to relay, we propose deploying an \ac{IRS} with $M$ reflecting elements, as in Fig. \ref{Butterfly_scheme}. 

There are two stages in \ac{PNC}. In the first stage, $\mathrm{S_{1}}$ and $\mathrm{S_{2}}$ broadcast their data to the relay and to destinations $\mathrm{D_{1}}$ and $\mathrm{D_{2}}$, respectively. Signals from $\mathrm{S_{1}}$ and $\mathrm{S_{2}}$ are assumed to arrive at the relay with symbol-level synchronization \cite{sykora2018wireless}. 

Let $P$ and $s_{i}$ be the  transmit power and the transmitted symbol of source $i$, respectively. Hence, the signal transmitted by source $i$ is $x_{i}=\alpha_{i}s_{i}$, where $\alpha_{i}=\sqrt{P/\sigma_{s,i}^{2}}$. The power of $s_{i}$ is $\mathbb{E}\left[|s_{i}|^{2}\right]=\sigma_{s,i}^{2}$. We assume \ac{BPSK} modulation at source nodes, but this also can be extended to \ac{QPSK}. The noise at the $j$th antenna of the relay  is assumed to be complex Gaussian with $\mu$ mean and $\sigma^{2}$ variance, i.e., $n_{j}\sim \mathcal{CN}(\mu,\,\sigma^{2})$. We consider Rayleigh fading channels between \ac{IRS} and relay, between sources and \ac{IRS}, between sources and relay, between $\mathrm{S}_{1}$ and $\mathrm{D}_{1}$, and between relay and $\mathrm{D}_{1}$ as $\mathbf{H^{ir}}\in\mathbb{C}^{\Nr\times M}$, $\mathbf{H^{ui}}\in\mathbb{C}^{M\times\Ns}$, $\mathbf{H^{ur}}\in\mathbb{C}^{\Nr\times\Ns}$, $h_{\mathrm{D}_{1}}^{\mathrm{S}_{1}}\in\mathbb{C}^{1\times 1}$ and $\mathbf{h}_{\mathrm{D}_{1}}^{r}\in\mathbb{C}^{1\times\Nr}$, respectively.
The received signal at the relay node can be written in a matrix form as
\begin{equation}
\mathbf{r}=\left(\mathbf{H^{ir}}\mathbf{\Theta} \mathbf{H^{ui}}+\mathbf{H^{ur}}\right)\mathbf{x}+\mathbf{n}\triangleq\mathbf{H}\mathbf{x}+\mathbf{n},
\label{relay_PNCc}
\end{equation}
where $\mathbf{\Theta}=\operatorname{diag}\left(\mathbf{v}\right)$ is the \ac{IRS} diagonal phase shift matrix with $\mathbf{v}=\left[e^{j\theta_{1}}, \ldots, e^{j\theta_{M}}\right]^{T}$ and $\theta_{m} \in \left[0,2\pi\right]$, $\mathbf{x}=\left[x_{1}, x_{2} \right]^{T}$ and $\mathbf{n}=\left[n_{1}, n_{2}, \ldots, n_{\Nr}\right]^{T}$ are the signal and noise vectors, respectively.

In the second stage, the relay computes an estimate of the XOR of the signals from $\mathrm{S_{1}}$ and $\mathrm{S_{2}}$.
Here, the XORed version is referred to as the network-coded form.
Then, the relay broadcasts the estimated network-coded signal to $\mathrm{D_{1}}$ and $\mathrm{D_{2}}$. The nodes $\mathrm{D_{1}}/\mathrm{D_{2}}$ can compute $x_{2}/x_{1}$ by XORing  $x_{1}/x_{2}$ and a network-coded signal from the relay. 


In naive \ac{NNC} rather than \ac{PNC}, the relay computes the estimates of $x_{1}$ and $x_{2}$ before creating a network-coded form $\x \triangleq x_{1} \oplus x_{2}$. However, such schemes are suboptimal since they do not consider that only $\x$ is needed at the relay rather than individual signals $x_{1}$ and $x_{2}$ \cite{sykora2018wireless,zhang2010physical}. 

Therefore, we propose \ac{PNC} approach where the relay computes an estimate of $\x$ without decoding $x_{1}$ and $x_{2}$ individually. In fact, it is more useful to get the estimate of $\x$ from $x_{1}+x_{2}$ and $x_{1}-x_{2}$, which can be calculated directly from $\mathbf{r}$ via matrix multiplication. 
Hence, the relay can get $\x$ almost at full rate \cite{wilson2010joint,zhang2010physical}. 

Based on this observation, the received signal in \eqref{relay_PNCc} can be rewritten as
\begin{equation}
\mathbf{r}=\left(\mathbf{H} \mathbf{D}^{-1}\right)(\mathbf{D} \mathbf{x})+\mathbf{n}=\mathbf{\widetilde{H}} \xtilde+\mathbf{n} \mbox{,}
\end{equation}
where \begin{equation}
    \mathbf{D}=2\mathbf{D}^{-1}=\left[\begin{array}{cc}1 & 1 \\ 1 & -1\end{array}\right]
\end{equation}
is the sum-difference matrix, and $\xtilde$ is
\begin{equation}
\xtilde=\left[\begin{array}{c}
\widetilde{x}_{1} \\
\widetilde{x}_{2}
\end{array}\right]\triangleq \mathbf{D}\,\mathbf{x}=\left[\begin{array}{c}
x_{1}+x_{2} \\
x_{1}-x_{2}
\end{array}\right]\mbox{.}
\end{equation}
The vector $\xtilde$ can be estimated from $\mathbf{r}$ using a linear operator as
\begin{align}
    \mathbf{y}=\mathbf{G}\mathbf{r} \mbox{,}
\end{align}
where $\mathbf{G}\in \mathbb{C}^{\Ns \times \Nr}$ is the beamforming matrix at the relay, designed in Section~\ref{sec:algorithm}. We consider that the relay can estimate the \ac{CSI}.


\section{Problem Formulation}\label{sec:problem_form}
In this section, we aim to design the optimal linear operator, i.e., the beamforming matrix $\mathbf{G}$ and the \ac{IRS} phase profile $\mathbf{\Theta}$ to minimize the \ac{MSE} for the recovery of $\xtilde$.  In this regard, the problem can be formulated as
\begin{subequations}\label{main_opt_problem}
\begin{alignat}{2}
&\underset{\boldsymbol{\Theta}, {\mathbf G}}{\operatorname{minimize}} \quad && \mathsf{MSE} \triangleq \mathbb{E}\left[\|\mathbf{y}(\boldsymbol{\Theta}, {\mathbf G})-\xtilde\|^{2}\right] \\
&\text {subject to} \quad && 0\leq\theta_{m}\leq2\pi, \ m \in{\{1,2,\ldots,M\}}.
\end{alignat}
\end{subequations}
The \ac{MSE} between the estimate $\mathbf{y}$ and target parameter $\xtilde$ can be computed as 
\begin{align}
    \mathsf{MSE} 
    &=\operatorname{tr}\left(\mathbb{E}\{\left(\mathbf{G}\mathbf{r}-\xtilde\right)\left(\mathbf{r}^{H}\mathbf{G}^{H}-\xtilde^{H}\right)\}\right) \nonumber \\
    &=\operatorname{tr}\left(\mathbf{G}\mathbb{E}\{\mathbf{r}\mathbf{r}^{H}\}\mathbf{G}^{H}-2\Re\{\mathbf{G}\mathbb{E}\{\mathbf{r}\xtilde^{H}\}\}+\mathbb{E}\{\xtilde\xtilde^{H}\}\right)\nonumber \\
    &=\operatorname{tr}\left(\mathbf{G}(P\mathbf{H}\mathbf{H}^{H}+\sigma^{2}\mathbf{I})\mathbf{G}^{H}-2\Re\{\mathbf{G}P\mathbf{H}\mathbf{D}^{H}\}+2P\mathbf{I}\right)\mbox{.}
    \label{MSE}
\end{align}

The joint optimization over the beamforming matrix and \ac{IRS} profile in \eqref{main_opt_problem} is challenging. In Section\,\ref{sec:algorithm}, we provide an efficient algorithm for the optimization problem. 
\section{Optimization Algorithm}
\label{sec:algorithm}
We propose an algorithm that optimizes the \ac{IRS} phases while fixing the beamforming matrix in \eqref{main_opt_problem}. In particular, for a fixed $\mathbf{\Theta}$ in \eqref{main_opt_problem}, the optimization problem is a quadratic unconstrained problem in $\mathbf{G}$ that can be solved in a closed-form as 
\begin{align}
    \mathbf{G}=(P\mathbf{H}^{H}\mathbf{H}+\sigma^{2}\mathbf{I})^{-1}P\mathbf{H}^{H} \mbox{,}
    \label{eq:MMSEG}
\end{align}
which is the \ac{MMSE} estimator for $\mathbf{G}$. To optimize the phases of \ac{IRS} for fixed $\mathbf{G}$, we propose the matrix lifting technique \cite{luo2010semidefinite}. First, the  objective function should be written  in terms of $\mathbf{v}$, i.e., $\mathsf{MSE}$ can be written as
\begin{multline}\label{MSE_v}
    \mathsf{MSE}\triangleq P\sum_{i^{\prime}=1}^{\Ns}\sum_{j{\prime}=1}^{\Nr}\left(\sum_{j=1}^{\Nr}\mathbf{g}_{j}^{H}\mathbf{g}_{j^{\prime}}\left(b^{ji^{\prime}}\right)^{H}\right)b^{j^{\prime}i^{\prime}}+2P\Ns   \\
    -2\,P\,\Re\left(\sum_{i=1}^{\Ns}\sum_{j=1}^{\Nr}b^{ji}\left(g_{1j}+(-1)^{i+1}g_{2j}\right)\right)+\sum_{i=1}^{\Ns} {{\sigma}_{i}}^{2} \mbox{,}
\end{multline}
where $b^{ji}=\boldsymbol{\phi}^{ji}\mathbf{v}+h_{ji}^{ur}$ and $\boldsymbol{\phi}^{ji}=\mathbf{h}^{\mathbf{ir}}_{(j)}\operatorname{diag}(\mathbf{h}^{\mathbf{ui}}_{i})$.

\noindent By substituting \eqref{eq:MMSEG} in  \eqref{MSE_v}, we can write the optimization problem \eqref{main_opt_problem} for a fixed  $\mathbf{G}$ as
\begin{subequations}\label{main_opt_problem_fixG}
\begin{alignat}{2}
&\underset{\mathbf{v}}{\operatorname{minimize}} \quad && \mathsf{MSE} \\
&\text {subject to} \quad && |\mathbf{v}(m)|^{2}=1, \ m \in\{1,2,\ldots,M\}\mbox{.}
\label{v_modulus_constraint}
\end{alignat}
\end{subequations}
The unit modulus constraint induces non-convexity. In this regard, we apply a matrix lifting technique \cite{luo2010semidefinite}. Now, the problem is tackled with the help of matrix $\mathbf{V}$, where
\begin{align}
    \mathbf{V}=\left[\begin{array}{c}\mathbf{v} \\ 1\end{array}\right]\left[\begin{array}{cc}\mathbf{v}^{H} & 1\end{array}\right]=\left[\begin{array}{cc}\mathbf{v}\mathbf{v}^{H}& \mathbf{v}\\\mathbf{v}^{H} & 1\end{array}\right]\mbox{.}
    \label{big_V}
\end{align}
For $\mathbf{V}$, we have the following constraints:  $\operatorname{rank}\left(\mathbf{V}\right)\leq1$, ${\mathbf{V}\geq0}$, and $\mathbf{V}(k,k)=1$, $\forall k\in\{1, \ldots ,M+1\}$. Let us rewrite \eqref{MSE_v} in terms of $\mathbf{V}$ and remove all the terms that do not depend on $\mathbf{v}$ as
\begin{align}\label{MSE_notdepend_v}
    \mathsf{MSE}&\propto \!\!\sum_{i^{\prime}=1}^{\Ns}\sum_{j{\prime}=1}^{\Nr}\sum_{j=1}^{\Nr}\!\!\operatorname{tr}\left(\mathbf{g}^{H}_{j}\mathbf{g}_{j^{\prime}}\!\!\left[\begin{array}{cc}{\boldsymbol{\phi}^{ji^{\prime}}}^{H}\boldsymbol{\phi}^{j^{\prime}i^{\prime}} & {\boldsymbol{\phi}^{ji^{\prime}}}^{H}h_{j^{\prime}i^{\prime}}^{ur} \\ h_{ji^{\prime}}^{*ur}\boldsymbol{\phi}^{j^{\prime}i^{\prime}} & 0\end{array}\right]\mathbf{V}\right) \nonumber\\ 
    &\phantom{\propto}-\sum_{i=1}^{\Ns}\sum_{j=1}^{\Nr}\operatorname{tr}\left(\left[\begin{array}{cc} 0 & \left(g_{1j}^{*}+(-1)^{i+1}g_{2j}^{*}\right){\boldsymbol{\phi}^{ji}}^{H} \\ 0 & 0\end{array}\right]\mathbf{V}\right) \nonumber\\  &\phantom{\propto}-\sum_{i=1}^{\Ns}\sum_{j=1}^{\Nr}\operatorname{tr}\left(\left[\begin{array}{cc} 0 & 0 \\ \left(g_{1j}+(-1)^{i+1}g_{2j}\right)\boldsymbol{\phi}^{ji} & 0\end{array}\right]\mathbf{V}\right) \nonumber\\
    &=\operatorname{tr}\left(\mathbf{A}\mathbf{V}\right) \mbox{,}
\end{align}
where $\mathbf{A}$ is a matrix representing the summation of all the matrices inside the trace function in \eqref{MSE_notdepend_v}.  
So, we have the following optimization problem
\begin{algorithm}[t!] 
  	\caption{Algorithm of Optimizing $\mathbf{V}$ for Fixed $\mathbf{G}$}
  	\label{alg.1} 
  	\begin{algorithmic} [1]
  		\STATE\textbf{Initialize}
  	    $\mathbf{V}_{\mathrm{iter}}$ as in \eqref{big_V} with an arbitrary vector $\mathbf{v}$ that satisfies \eqref{v_modulus_constraint} and set $\mathrm{iter}=1$
  		\REPEAT
  		\STATE \textbf{Step 1:} Set $\mathbf{V}^{\prime}:=\mathbf{V}_{\mathrm{iter}}$
  		\STATE \textbf{Step 2:}
  		 Solve convex problem \eqref{final_opt_problem}, whose output is $\mathbf{V}$.
  		\STATE \textbf{Step 3:} Update $\mathbf{V}_{\mathrm{iter}}:=\mathbf{V}$
  		\STATE \textbf{Step 4:} Set $\mathrm{iter}=\mathrm{iter}+1$
  		 \UNTIL{$\left||\mathbf{V}_{\mathrm{iter}}-\mathbf{V}^{\prime}\right||^2_{F}\leq\delta$ or $\mathrm{iter}>\mathrm{Iter}_{\max}$}
  		\STATE $\mathbf{V}:=\mathbf{V}_{\mathrm{iter}}$
  		\STATE \textbf{Output} $\mathbf{V}$
  	\end{algorithmic}
  \end{algorithm}
\begin{subequations}\label{main_opt_problem_fixG_MV}
\begin{alignat}{2}
&\underset{\mathbf{V}\geq0}{\operatorname{minimize}} \quad &&  \operatorname{tr}\left(\mathbf{A}\mathbf{V}\right) \\
&\text {subject to} \quad && \mathbf{V}(k,k)=1, \ k \in\{1,2,\ldots,M+1\} \\
& \quad && \operatorname{rank}\left(\mathbf{V}\right)\leq1 \mbox{.}
\end{alignat}
\end{subequations}
The constraint $\operatorname{rank}\left(\mathbf{V}\right)\leq1$ is non-convex and equivalent to 
\begin{align}
    \operatorname{tr}\left(\mathbf{V}\right)-\beta_{1}\left(\mathbf{V}\right)=0,
    \label{eq_rank}
\end{align}
where $\beta_{1}\left(\mathbf{V}\right)$ is the largest singular value of the matrix $\mathbf{V}$. Also, for $\mathbf{V}\geq0$, i.e., semidefinite, the left side of \eqref{eq_rank} is $0$ when $\operatorname{rank}\left(\mathbf{V}\right)\leq1$ is satisfied; otherwise, it is greater than zero. The optimization problem can be written as
\begin{subequations}\label{main_opt_problem_fixG_MV_gamma}
\begin{alignat}{2}
&\underset{\mathbf{V}\geq0}{\operatorname{minimize}} \quad && \left(\operatorname{tr}\left(\mathbf{A}\mathbf{V}\right)+\gamma\left(\operatorname{tr}\left(\mathbf{V}\right)-\beta_{1}\left(\mathbf{V}\right)\right)\right) \label{eq:objectivelagrange} \\
&\text {subject to} \quad && \mathbf{V}(k,k)=1, \ k \in\{1,2,\ldots,M+1\} \mbox{,}
\end{alignat}
\end{subequations}
where $\gamma\geq0$ is a fixed weight.  Since $\beta_{1}\left(\mathbf{V}\right)$  and $\operatorname{tr}\left(\mathbf{A}\mathbf{V}\right)+\gamma\,\operatorname{tr}\left(\mathbf{V}\right)$ are convex functions in $\mathbf{V}$, \eqref{main_opt_problem_fixG_MV_gamma} can be represented as a \ac{DC} problem. \Ac{CCP} can efficiently obtain a local optimal solution for \ac{DC} problems \cite{tao2016content}.
In this regard, we linearize $-\gamma\,\beta_{1}(\mathbf{V})$, which induces non-convexity in \eqref{main_opt_problem_fixG_MV_gamma}, around a point $\mathbf{V}=\mathbf{V}^{\prime}$ using the following upper bound \cite{jiang2019over}
\begin{align}
    -\gamma\,\beta_{1}\left(\mathbf{V}\right)\leq-\gamma\, \operatorname{tr}\left(\mathbf{V}\mathbf{u_{1}}\left(\mathbf{V}^{\prime}\right)\mathbf{u_{1}}\left(\mathbf{V}^{\prime}\right)^{H}\right)\mbox{,}
    \label{penalty_linearrizing}
\end{align}
where $\mathbf{u_{1}}\left(\mathbf{V}^{\prime}\right)$ is the eigenvector of the  matrix $\mathbf{V}^{\prime}$ corresponding to its largest eigenvalue. The equality condition in \eqref{penalty_linearrizing} is satisfied when $\mathbf{V}=\mathbf{V}^{\prime}$. We get the following convex optimization problem, which can be  solved by {\tt cvx}:
\begin{subequations}\label{final_opt_problem}
\begin{alignat}{2}
&\underset{\mathbf{V}\geq0}{\operatorname{min}} \quad && \operatorname{tr}\left(\mathbf{A}\mathbf{V}\right)+\gamma\left(\operatorname{tr}\left(\mathbf{V}\right)-\operatorname{tr}\left(\mathbf{V}\mathbf{u_{1}}\left(\mathbf{V}^{\prime}\right)\mathbf{u_{1}}\left(\mathbf{V}^{\prime}\right)^{H}\right)\right) \\
&\text {s.t.} \quad && \mathbf{V}(k,k)=1, \ k \in\{1,2,\ldots,M+1\} \mbox{.}
\end{alignat}
\end{subequations}
Appropriate choice of the penalty parameter $\gamma$ in \eqref{final_opt_problem} can be found via simple bisection and  remains static throughout Alg.~\ref{alg.1}. An  iterative algorithm that alternately optimizes the phases of IRS and  beamforming matrix $\mathbf{G}$ is given in Alg.~\ref{alg.2}.
\begin{algorithm}[t!]
  	\caption{Iterative Algorithm for \ac{MMSE} Optimization} 
  	\label{alg.2} 
  	\begin{algorithmic} [1]
  		\STATE\textbf{Initialize}
  	     $\mathbf{v}^{1}$  as an arbitrary vector that satisfies \eqref{v_modulus_constraint} and set $k=0$
  		\REPEAT
  		\STATE \textbf{Step 1:} Set $k=k+1$
  		\STATE \textbf{Step 2:} Update $\mathbf{G}^{k}$ using \eqref{eq:MMSEG}, where diagonal elements of phase matrix $\mathbf{\Theta}$ are the elements of vector $\mathbf{v}^{k}$
  		\STATE \textbf{Step 3:} Run Alg.~\ref{alg.1}, where $\mathbf{G}=\mathbf{G}^{k}$, and update $\mathbf{v}^{k+1}$ with the first $M$ elements in $\mathbf{v}_{M+1}$ 
  		 \UNTIL{$\left||\mathbf{v}_{k+1}-\mathbf{v}^{k}\right||^2+\left||\mathbf{G}_{k+1}-\mathbf{G}^{k}\right||^2_{F} <\delta$}
  	\end{algorithmic}
  \end{algorithm}
\vspace{-0.7cm}
\section{Optimal Detector and Error Performance}
\subsection{Optimal Detector}
The likelihood estimator can be considered to estimate the network coding form, i.e., $\x$, from $\mathbf{y}$. In this regard, the \ac{LLR} can be written, ignoring the noise dependencies in $y_{1}$ and $y_{2}$, as

\begin{align}
\label{LLR}
&\mathcal{L}\left(\x | y_{1} y_{2}\right)=\log\left(\frac{\mathbb{P}\{y_{1} y_{2} | \x=1\}}{\mathbb{P}\{y_{1} y_{2} | \x=-1\}}\right) \nonumber \\
&=\log\left(\frac{\mathbb{P}\{y_{1} | \widetilde{x}_{1}=0\}\left[\mathbb{P}\{y_{2} | \widetilde{x}_{2}=2\}+\mathbb{P}\{y_{2} | \widetilde{x}_{2}=-2\}\right]}{\left[\mathbb{P}\{y_{1} | \widetilde{x}_{1}=2\}+\mathbb{P}\{y_{1} | \widetilde{x}_{1}=-2\}\right] \mathbb{P}\{y_{2} | \widetilde{x}_{2}=0\}}\right) \nonumber \\
&=2\left({\frac{1}{\sigma_{1}^{2}}-\frac{1}{ \sigma_{2}^{2}}}\right)+\log\left(\frac{\cosh \left(2 y_{2} / \sigma_{2}^{2}\right)}{\cosh \left(2 y_{1} /\sigma_{1}^{2}\right)}\right) \mbox{,}
\end{align}
where $ \sigma^{2}_{i}\triangleq\{\mathbf{G}\mathbf{G}^{H}\}_{i,i} \sigma^{2}$ is the noise variance on the $i$th stream after the linear operator. Let us first define the estimation of $\x$ at the relay and $\mathrm{D_{1}}$ as $\xrelay$ and $\xDone$, respectively. The corresponding decision rules are then
\begin{align}
    \xrelay&=
\begin{cases}
1 & \text { when } \mathcal{L}\left(\x| y_{1} y_{2}\right) \geq 0 \\
-1 & \text { when } \mathcal{L}\left(\x| y_{1} y_{2}\right)<0
\end{cases}\mbox{,} \\
\xDone&=
\begin{cases}
1 & \text { when } \mathcal{L}_{\mathrm{D}_{1}}^{r}\left(\xrelay| y_{\mathrm{D}_{1}}^{r}\right) \geq 0 \\
-1 & \text { when } \mathcal{L}_{\mathrm{D}_{1}}^{r}\left(\xrelay| y_{\mathrm{D}_{1}}^{r}\right)<0
\end{cases}\mbox{,}
\end{align}
where $\mathcal{L}_{\mathrm{D}_{1}}^{r}=2y_{\mathrm{D}_{1}}^{r}/\sigma_{r,\mathrm{D}_{1}}^{2}$  is the likelihood detector, and $y_{\mathrm{D}_{1}}^{r}$ is the received signal at $\mathrm{D}_{1}$ from relay. After the channel inversion, the variance of the noise at $\mathrm{D}_{1}$ is $\sigma_{r,\mathrm{D}_{1}}^{2}=\sigma^{2}/\lVert\mathbf{h}_{\mathrm{D}_{1}}^{r}\rVert^{2}$. 
After that, the relay broadcasts $\xrelay$ to the destinations. $\mathrm{D_{1}}$ receives $\xDone$ and $\widehat{x}_{1}$, which is estimated similarly to $\xDone$ with likelihood detector $\mathcal{L}_{\mathrm{D}_{1}}^{\mathrm{S}_{1}}=2y_{\mathrm{D}_{1}}^{\mathrm{S}_{1}}/\sigma_{\mathrm{S}_{1},\mathrm{D}_{1}}^{2}$, received signal $y_{\mathrm{D}_{1}}^{\mathrm{S}_{1}}$, and noise variance after channel inversion  $\sigma_{\mathrm{S}_{1},\mathrm{D}_{1}}^{2}=\sigma^{2}/\lvert h_{\mathrm{D}_{1}}^{\mathrm{S}_{1}}\rvert^{2}$. 
 Then, $\widehat{x}_{2}$ is obtained at $\mathrm{D}_{1}$ by XORing $\xDone$ and $\widehat{x}_{1}$. Similarly, $\mathrm{D_{2}}$  receives $\widehat{x}_{2}$ and $\xDtwo$, and obtains $\widehat{x}_{1}$ by XORing the two values. 
\vspace{-0.1cm}
\subsection{BER Analysis}
\label{sec_ber_analysis}
Before calculating instantaneous theoretical \ac{BER} at $\mathrm{D}_{1}$, i.e., $\BERD\triangleq \mathbb{P}\left(\widehat{x}_{2}\neq x_{2}\right)$, we can find the instantaneous \acp{BER} separately for the links from sources to relay, from $\mathrm{S}_{1}$ to $\mathrm{D}_{1}$, and from relay to $\mathrm{D_{1}}$. Let us define the set $\mathbb{X}\triangleq\{-2,2\}$, which contains all the values of $\widetilde{x}_{2}$ and $\widetilde{x}_{1}$ such that $\x=1$ and $\x=-1$, respectively. To obtain $\xtilde$  at the relay from $y_{1}$ and $y_{2}$, we use the \ac{LLR} in \eqref{LLR}. To derive $\BERrelay\triangleq \mathbb{P}\left(\xrelay\neq\x \right)$ at the relay, we need to further rewrite \eqref{LLR} as
\begin{align}\label{extended_LLR}
\mathcal{L}\left(\x | y_{1} y_{2}\right)&=
\sum_{i=1}^{\Ns}\left(-1\right)^{i}\left[\log\left(\sum_{\widetilde{x}_{i}\in\mathbb{X}}e^{-\frac{(y_{i}-\widetilde{x}_{i})^{2}}{2\sigma_{i}^{2}}}\right)+\frac{y_{i}^{2}}{2\sigma_{i}^{2}}\right] \mbox{.}
\end{align}
Now using the soft minimum approximation $\log\left(\sum_{j}\exp{\left(-Z_{j}\right)}\right)\approx -\min_{j}(Z_j)$ from \cite{calafiore2020universal}, we can approximate the \ac{LLR} in \eqref{extended_LLR} as
\begin{align}
\bar{\mathcal{L}}\left(\x | y_{1} y_{2}\right)&=\sum_{i=1}^{\Ns}\left(-1\right)^{i}\left[-\min_{\widetilde{x}_{i}\in \mathbb{X}}\left(\frac{(y_{i}-\widetilde{x}_{i})^{2}}{2\sigma_{i}^{2}}\right)+\frac{y_{i}^{2}}{2\sigma_{i}^{2}}\right]\nonumber \\
&=\sum_{i=1}^{\Ns}\left(-1\right)^{i+1}\min_{\widetilde{x}_{i}\in\mathbb{X}}\left(\frac{-2y_{i}\widetilde{x}_{i}+\widetilde{x}_{i}^{2}}{2\sigma_{i}^{2}}\right) \mbox{.}
\end{align}
Hence, instantaneous $\BERrelay$ can be approximated as 
\begin{align}
    \BERrelay&\approx\mathbb{P}\{\x=1\} \mathbb{P}\{\bar{\mathcal{L}}<0|\x=1\}\nonumber \\
    &\phantom{\approx}+\mathbb{P}\{\x=-1\}\mathbb{P}\{\bar{\mathcal{L}}\geq0|\x=-1\} \nonumber \\
    &=\frac{1}{2}\left[Q\left(\frac{-2}{\sigma_{2}}\right)+Q\left(\frac{-2}{\sigma_{1}}\right)\right] Q\left(\frac{\sqrt{1+\frac{\sigma_{2}^{2}}{\sigma_{1}^{2}}}}{\sigma_{2}}\right) \nonumber \\
    &\phantom{\approx}+Q\!\!\left(\!\frac{2}{\sigma_{2}}\!\right)\!Q\!\!\left(\!\!\frac{-3+\frac{\sigma_{2}^{2}}{\sigma_{1}^{2}}}{\sigma_{2}\sqrt{1+\frac{\sigma_{2}^{2}}{\sigma_{1}^{2}}}}\!\!\right)+Q\!\!\left(\!\frac{2}{\sigma_{1}}\!\right) \!Q\!\!\left(\!\frac{-3+\frac{\sigma_{1}^{2}}{\sigma_{2}^{2}}}{\sigma_{1}\sqrt{1+\frac{\sigma_{1}^{2}}{\sigma_{2}^{2}}}}\!\right)
    \!\!\mbox{,}
\end{align}
where $Q(\cdot)$ is the Q-function \cite{john2008digital}. Using the channel inversion precoder and likelihood detectors \cite{john2008digital}, the \acp{BER} at $\mathrm{D}_{1}$ for a given channel realization are  $\BERSD\triangleq \mathbb{P}\left(\widehat{x}_{1}\neq x_{1}\right)=Q\left(1/\sigma_{\mathrm{S}_{1},\mathrm{D}_{1}}\right)$ and $\BERRD\triangleq \mathbb{P}\left(\xDone\neq \xrelay\right)=Q\left(1/\sigma_{r,\mathrm{D}_{1}}\right)$. Now, $\BERD$ can be derived as
\begin{align}
\label{eq:BERD1}
    \BERD&=\BERSD\left(1-\BERRD\right)\left(1-\BERrelay\right)+\BERSD\BERRD\BERrelay +\left(1\!-\!\BERSD\right) \nonumber \\
    &\phantom{=}\times \BERRD\left(1-\BERrelay\right)+\left(1-\BERSD\right)\left(1-\BERRD\right)\BERrelay \mbox{.}
\end{align}
Note that the event $\left(\widehat{x}_{2}\neq x_{2}\right)$ at $\mathrm{D}_{1}$ occurs when the number of links with error is odd. For example, $\BERSD\left(1-\BERRD\right)\left(1-\BERrelay\right)$ considers that the estimated value of $x_{1}$ at $\mathrm{D}_{1}$ is wrong while $\xrelay$ and $\xDone$ are correct. 

\section{Simulation Results}\label{sec_simulation_results}
In this section, we demonstrate the advantages of using \ac{IRS} and \ac{PNC} in terms of the \ac{BER}. The  channel coefficients are generated as standard complex Gaussian random variables, and the \ac{SNR} is defined as $\mathrm{SNR}\triangleq P/\sigma^2$. In all figures, solid and dashed lines represent theoretical and Monte-Carlo simulation results, respectively.

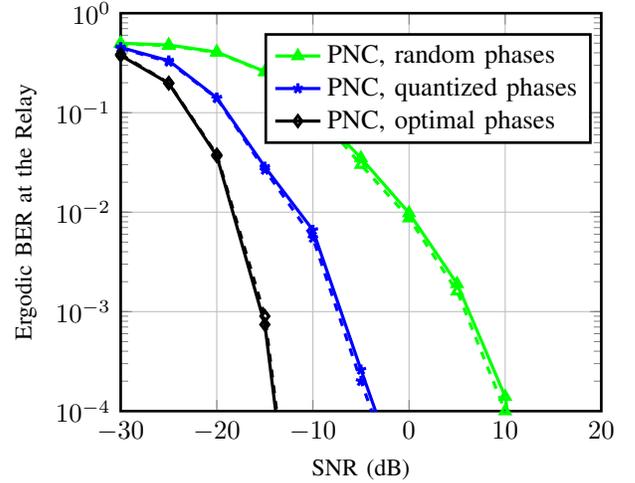
\begin{figure}
\centering
	\pgfplotsset{every axis/.append style={
		line width=1pt,
		legend style={font=\normalsize, at={(0.98,0.95)}},legend cell align=left},
} %
\pgfplotsset{compat=1.13}
	\begin{tikzpicture}
\begin{semilogyaxis}[
xlabel near ticks,
ylabel near ticks,
grid=major,
xlabel={\ac{SNR} (dB)},
ylabel={Ergodic \ac{BER} at the Relay},
width=0.9\linewidth,
yticklabel style={/pgf/number format},
legend entries={\ac{PNC}\mbox{,} random phases, ,\ac{PNC}\mbox{,} quantized phases, , \ac{PNC}\mbox{,} optimal phases, ,},
	xmin=-30, xmax=20,
	ymin=1e-4, ymax=1,
xlabel style={font=\small},
ylabel style={font=\small},
]
\addplot[green,mark=triangle,very thick,mark options=solid] table {Figures/BER_vs_SNR_R_unnormalized/Data_PNC_random_th_relay.dat};
\addplot[green,mark=triangle,dashed,very thick,mark options=solid] table {Figures/BER_vs_SNR_R_unnormalized/Data_PNC_random_relay.dat};
\addplot[blue,mark=star,very thick,mark options=solid] table {Figures/BER_vs_SNR_R_unnormalized/Data_PNC_quant_th_relay.dat};
\addplot[blue,mark=star,dashed,very thick,mark options=solid] table {Figures/BER_vs_SNR_R_unnormalized/Data_PNC_quant_relay.dat};
\addplot[black,mark=diamond,very thick,mark options=solid] table {Figures/BER_vs_SNR_R_unnormalized/Data_PNC_optimal_th_relay.dat};
\addplot[black,mark=diamond,dashed,very thick,mark options=solid] table {Figures/BER_vs_SNR_R_unnormalized/Data_PNC_optimal_monte_relay.dat};
\end{semilogyaxis}
\end{tikzpicture}
	\caption{Ergodic \ac{BER} at the relay, $\BERrelay$, as a function of SNR, for $M=32$.}
\label{BER_relay}
\vspace{-0.3cm}
\end{figure}
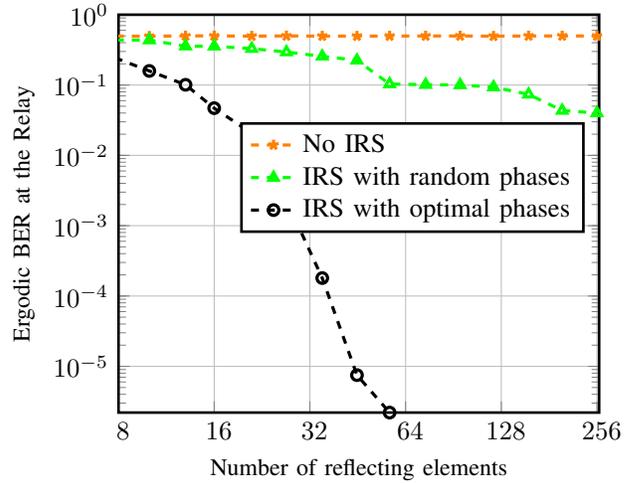
\begin{figure}
\centering
	\pgfplotsset{every axis/.append style={
		line width=1pt,
		legend style={font=\normalsize, at={(0.97,0.73)}},legend cell align=left},
} %
\pgfplotsset{
  log x ticks with fixed point/.style={
      xticklabel={
        \pgfkeys{/pgf/fpu=true}
        \pgfmathparse{exp(\tick)}%
        \pgfmathprintnumber[fixed relative, precision=3]{\pgfmathresult}
        \pgfkeys{/pgf/fpu=false}
      }
  },
  log y ticks with fixed point/.style={
      yticklabel={
        \pgfkeys{/pgf/fpu=true}
        \pgfmathparse{exp(\tick)}%
        \pgfmathprintnumber[fixed relative, precision=3]{\pgfmathresult}
        \pgfkeys{/pgf/fpu=false}
      }
  }
}

\pgfplotsset{compat=1.13}
	\begin{tikzpicture}
\begin{loglogaxis}[
xlabel near ticks,
ylabel near ticks,
grid=major,
xlabel={Number of reflecting elements},
ylabel={Ergodic \ac{BER} at the Relay},
width=0.9\linewidth,
xtick={8,16,32,64,128,256},
log x ticks with fixed point,
legend entries={No \ac{IRS},\ac{IRS} with random phases 
	,\ac{IRS} with optimal phases,},
	xmin=8, xmax=260,
	ymin=2.2e-6, ymax=1,
ylabel style={font=\small},
xlabel style={font=\small},
]
\addplot[orange,mark=star,dashed,very thick,mark options=solid] table {Figures/BER_vs_M_unnormalized/Data_PNC_no_IRS.dat};
\addplot[green,mark=triangle,dashed,very thick,mark options=solid] table {Figures/BER_vs_M_unnormalized/Data_PNC_random.dat};
\addplot[black,mark=o,dashed,very thick,mark options=solid] table {Figures/BER_vs_M_unnormalized/Data_PNC_optimal.dat};


\end{loglogaxis}
\end{tikzpicture}\vspace{-4mm}
	\caption{Ergodic \ac{BER} at the relay, $\BERrelay$, versus the number of reflecting elements at \ac{IRS}, for $\mathrm{SNR}=-15$~dB.}
	\label{BER_th_num_rel}
	\vspace{-0.35cm}
\end{figure}
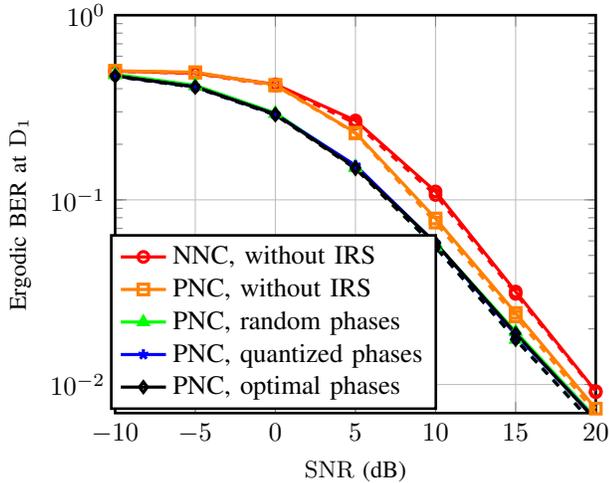
\begin{figure}
\centering
 	\pgfplotsset{every axis/.append style={
		line width=1pt,
		legend style={font=\normalsize, at={(0.67,0.45)}},legend cell align=left},
} %
\pgfplotsset{compat=1.13}
	\begin{tikzpicture}
\begin{semilogyaxis}[
xlabel near ticks,
ylabel near ticks,
grid=major,
xlabel={$\mathrm{SNR}$ (dB)},
ylabel={Ergodic \ac{BER} at $\mathrm{D}_{1}$},
width=0.9\linewidth,
yticklabel style={/pgf/number format},
legend entries={\ac{NNC}\mbox{,} without \ac{IRS}, ,\ac{PNC}\mbox{,} without \ac{IRS}, ,\ac{PNC}\mbox{,} random phases, ,\ac{PNC}\mbox{,} quantized phases, , \ac{PNC}\mbox{,} optimal phases, ,},
	xmin=-10, xmax=20,
	ymin=0.7e-2, ymax=1,
ylabel style={font=\small},
xlabel style={font=\small},
]
\addplot[red,mark=o,very thick,mark options=solid] table {Figures/BER_vs_SNR_D1_unnormalized/Data_NNC_no_IRS_th.dat};
\addplot[red,mark=o,dashed, very thick,mark options=solid] table {Figures/BER_vs_SNR_D1_unnormalized/Data_NNC_no_IRS_monte.dat};
\addplot[orange,mark=square,very thick,mark options=solid] table {Figures/BER_vs_SNR_D1_unnormalized/Data_PNC_no_IRS_th.dat}; 
\addplot[orange,mark=square,dashed, very thick,mark options=solid] table {Figures/BER_vs_SNR_D1_unnormalized/Data_PNC_no_IRS.dat}; 
\addplot[green,mark=triangle,very thick,mark options=solid] table {Figures/BER_vs_SNR_D1_unnormalized/Data_PNC_random_th.dat};
\addplot[green,mark=triangle,dashed,very thick,mark options=solid] table {Figures/BER_vs_SNR_D1_unnormalized/Data_PNC_random.dat};
\addplot[blue,mark=star,very thick,mark options=solid] table {Figures/BER_vs_SNR_D1_unnormalized/Data_PNC_quant_th.dat};
\addplot[blue,mark=star,dashed,very thick,mark options=solid] table {Figures/BER_vs_SNR_D1_unnormalized/Data_PNC_quant.dat};
\addplot[black,mark=diamond,very thick,mark options=solid] table {Figures/BER_vs_SNR_D1_unnormalized/Data_PNC_optimal_th.dat};
\addplot[black,mark=diamond,dashed,very thick,mark options=solid] table {Figures/BER_vs_SNR_D1_unnormalized/Data_PNC_optimal_monte.dat};
\end{semilogyaxis}
\end{tikzpicture}
	\caption{Ergodic \ac{BER} at $\mathrm{D_{1}}$, $\BERD$, as a function of $\mathrm{SNR}$, for $M=32$.}
\label{BER_user3_x2}
\vspace{-0.35cm}
\end{figure}
 To evaluate the impact of the \ac{IRS} phase design, the \ac{BER} at the relay, $\BERrelay$, versus \ac{SNR} is depicted in Fig. \ref{BER_relay} for various phase designs, i.e., \textit{i) optimal:} the phases are designed according to the proposed scheme in Section~\ref{sec:algorithm}; \textit{ii) quantized:} the optimal phases are quantized to two levels ($0$ or $\pi$); \textit{iii) random:} the phases are selected uniformly at random from $0$ to $2\, \pi$. It is clear that the Monte-Carlo simulations coincide with the theoretical results. Also,  optimizing the \ac{IRS} phases can significantly decrease the \ac{BER}. For a target \ac{BER} of $10^{-2}$, the \ac{SNR} gain compared to random phases is about $18$~dB. 
 
The impact of the number of \ac{IRS} elements on the \ac{BER} at the relay in shown in Fig. \ref{BER_th_num_rel}. Two phase profiles are considered (optimal and random) along with the case without \ac{IRS}.  For random phases, $\BERD$ decreases slowly, while the proposed phase design allows the \ac{BER} to rapidly decrease with $M$. 
 
In Fig. \ref{BER_user3_x2}, we evaluate the impact of two network coding schemes (\ac{PNC} and \ac{NNC}) and two architectures (with and without \ac{IRS}) on the \ac{BER} at the destination node, $\mathrm{D_{1}}$. We can see that the proposed scheme, i.e., \ac{PNC} combined with \ac{IRS}, outperforms the schemes that consider \ac{PNC} and \ac{NNC} without \ac{IRS}. The \ac{PNC} with random and optimal phases have $\BERD=10^{-2}$ at $\text{\ac{SNR}}=17.85$~dB and $\text{\ac{SNR}}=17.8$~dB, respectively. Surprisingly, in contrast to the error experienced at the relay in Fig. \ref{BER_relay}, the gap between the \ac{PNC} approaches with different \ac{IRS} designs is negligible. This is attributed to the fact that the \ac{BER} for the direct link between $\mathrm{S}_{1}$ and $\mathrm{D}_{1}$, $\BERSD$,  dominates the effect of $\BERrelay$, while computing the \ac{BER} at the relay in \eqref{eq:BERD1}. Therefore, future works can consider adding  two \acp{IRS} to enhance the direct links from the sources to destinations, i.e., $\mathrm{S}_{1}-\mathrm{D}_{1}$ and $\mathrm{S}_{2}-\mathrm{D}_{2}$.  




\vspace{-0.1cm}
\section{Conclusion}\label{sec_conclusion}
We proposed an \ac{IRS}-aided \ac{PNC} to improve the wireless network throughput and \ac{BER} performance. The main contribution is our novel design with IRS that estimates the XOR value of two symbols over-the-air with optimal \ac{IRS} phases to minimize the estimation error. Also, we derived analytical expressions for the \ac{BER}. The numerical results show that jointly optimizing the IRS phases and  beamforming matrix at the relay offers better performance in terms of the \ac{BER}. For instance, the \ac{BER} at the relay in a $32$-element \ac{IRS}-assisted environment is three orders of magnitudes less than that without \acp{IRS}. For a target \ac{BER} of $10^{-2}$, we can achieve around $2$~dB performance gain in \ac{SNR} at destinations compared to naive network coding without \ac{IRS}. As future research directions, our proposed approach can be implemented and analyzed on more complicated network structures with multiple jointly-optimized \acp{IRS}. 
\vspace{-0.25cm}

\vspace{-0.05cm}

\bibliographystyle{IEEEtran}
\bibliography{ref.bib}

\end{document}